\documentstyle[12pt]{article}
\def \be{\begin{equation}}
\def \ee{\end{equation}}
\def \ba{\begin{array}{l}}
\def \ea{\end{array}}

\def \d{\delta}

\def \z{\zeta}

\def \ol{\overline}

\def \la{\langle}
\def \ra{\rangle}
\def \fr{\frac}
\def \2{\frac{1}{2}}
\def \4{\frac{1}{4}}
\def \lb{\label}
\newcommand{\sectio}[1]{\section{#1}\setcounter{equation}{0}}

\begin{document}
\frenchspacing
\setlength{\parskip}{2mm}

\vspace*{25mm}

\begin{center}

{\Large \bf The Wandering Exponent of a One-Dimensional Directed Polymer
in a Random Potential with Finite Correlation Radius}
\vskip .4in
S. E. Korshunov and Vik. S. Dotsenko
\vskip .2in
L. D. Landau Institute for Theoretical Physics,\\
Russian Academy of Sciences,\\
Kosygina 2, 117940 Moscow, Russia\\
~\\
~\\
\today

\end{center}
\vskip 1in

\begin{abstract}
We consider a one-dimensional directed polymer in a random potential which is
characterized by the Gaussian statistics with the finite size local
correlations.
It is shown that the well-known Kardar's solution obtained originally
for a directed polymer with $\d$-correlated random potential
can be applied for the description of the present system
only in the high-temperature limit.
For the low temperature limit we have obtained the new solution which is
described by the one-step replica symmetry breaking.
For the mean square deviation of the directed polymer of the linear size $L$
it provides the usual scaling $\ol{x^{2}} \sim a L^{2\z}$ with the wandering
exponent $\z = 2/3$ and the temperature-independent prefactor.
\end{abstract}
\newpage

\sectio{Introduction}

In a wide variety of physical systems one is interested in the behaviour of a
fluctuating linear object (with finite line tension)
interacting with a quenched random potential.
The object under consideration may be a dislocation in a crystal, a domain
wall in a two-dimensional magnet, a vortex line in a superconductor, a
fluxon line in an extended Josephson junction and so on, but
following   Ref. \cite{KZ} this class of problems
is traditionally discussed in terms of a directed polymer in
random media or symply "directed polymer".

Quite naturally the best understanding has been achieved for the simplest
one-dimensional case when the displacements of a directed polymer
can occur only in one direction.
In such case a directed polymer in continuous approximation can be described
by the Hamiltonian

\begin{equation}
H[x(t),v] = \int_{0}^{L} dt \, \left\{\frac{J}{2}\left(\frac{dx}{dt}\right)^2
+v[x(t),t]\right\}                                           \label{a1}
\end{equation}
where $J$ is the linear tension,
$t$ is the longitudal coordinate ($0 \leq t \leq L$) and
$x(t)$ is the transverse displacement of a polymer with respect
to a straight line.
The simplest (or maybe one should better say the most easily treatable)
assumption on the distribution of random potential $v(x,t)$
consists in taking it to be Gaussian with

\begin{equation}
\overline{v(x,t)}=0;~~~~{\overline{v(x,t)v(x',t')}}=2V(x-x')\delta (t-t').
                                                              \label{a2}
\end{equation}
Here and further on an overbar denotes the average over the realizations of
quenched random potential.

Let us assume that at $t=0$ the position of a polymer is fixed:
$x(0) \equiv 0$. Then the quantity of interest is the typical deviation
of the polymer "trajectory"  $x(t)$ from the origine.
More precisely, one would like to know the dependence on $L$
of the average square deviation of the polymer at the ending point $x=L$,
which in the limit $L\to\infty$ is expected to be described by the simple
scaling:

\begin{equation}
\lb{ab3}
\ol{\la x^{2}(L)\ra} \sim a L^{2\z}
\end{equation}
where angular brackets denote the average over thermal fluctuations,
and $\z$ is the so-called wandering exponent.

In  the  absence  of  the  random  potential  the situation is trivial and the
wandering  exponent  $\z$  is  equal  to  $1/2$.   In this case the trajectory
deviates from the origine only due to the thermal fluctuations, the  prefactor
$a$  in  the  scaling  law  (\ref{ab3}) is proportional to the
temperature, so at zero temperature $\la x^{2}(L)\ra = 0$.

In the presence of a quenched random potential the situation is getting much
more complicated.   Now besides  the thermal  fluctuations, the  trajectory is
pushed away  from the  origine also  due to  the randomness  in the background
potential landscape, so that the scaling law (\ref{ab3}) could be governed  by
a new non-trivial  wandering exponent.  Moreover,  since for a  generic random
potential the ground state trajectory of the Hamiltonian (\ref{a1})  typically
drifts away from the origine, the scaling law (\ref{ab3}) can be expected to
hold also in the zero temperature limit.

It is  widely believed  that  for a  whole class of {\it locally  correlated}
random  potentials  (such  that  the  function $V(x)$ in (\ref{a2}) is quickly
decaying for $|x|  \to \infty$) the  wandering exponent $\z$ is universal and
equal to 2/3.
This conclusion is based on the finite temperature exact results of Refs.
\cite{HHF,K}
and has been also confirmed by zero temperature numerical simulations
of the discrete version of the directed polymer problem \cite{KZ}
(see also Ref. \cite{HHZ} for later references).
However, both the calculation based on the reduction to damped Burgers'
equation with conservative random force \cite{HHF} and Bethe ansatz
calculation in terms of replica representation \cite{K} are valid only for
the case of stricktly $\delta$-functional correlations of random potential.

In this work we present an attempt to generalize the second
of these solutions to a more physical
situation when correlations of random potential have finite correlation
radius (in transverse direction).

In Sec. 2 we review some details of the approach developed by Kardar in Ref.
\cite{K} for $\delta$-functional correlations of random potential and use it
to find the temperature dependence of a prefactor in Eq. (\ref{ab3})
which turns out to be of the form
\begin{equation}
a\propto T^{-2/3}                                             \label{ab3b}
\end{equation}

In Sec. 3 the applicability of Kardar's solution \cite{K} for the
approximate description of the system with the finite radius $r$ of the random
potential correlations is discussed.
We demonstrate that it can be used only in the high
temperature limit $T\gg T_0\propto r^{2/3}$, whereas at low temperatures the
solution has to have essentially different structure which in principle could
lead to the change in wandering exponent.
The  impossibility  to  apply  the  description  with the help of the Kardar's
solution at arbitarily low temperatures follows already from Eq. (\ref{ab3b}).
One can expect that if in the low temperature limit the typical trajectory goes
away  from  the  origine  its  drift should  be  determined by
the quenched random potential and not by the effects of the thermal
fluctuations.  It  means that in the system with reasonable short-scale
{\em regularization} the divergence of the prefactor suggested by
Eq. (\ref{ab3b}) at low temperatures can be expected to saturate.

In Sec. 4 the low temperature solution of the regularized problem is found for
the particular choice of the random potential correlation function $V(x)$.
The form of this solution can be described in terms
of the effective one-step replica symmetry breaking ansatz.
In this case one recovers the scaling law (\ref{ab3})
with the same wandering exponent $\z = 2/3$ but
with a {\it temperature-independent} prefactor.

\sectio{The solution of the unregularized problem}
\subsection{The relation between exponents}

The idea of the Kardar's approach \cite{K} is based on the indirect calculation
of the wandering exponent $\z$ by analyzing the scaling of the typical sample
to sample {\it fluctuations} of the free energy.
Suppose that the typical fluctuations of the free energy
(produced by the random potential) scale as

\begin{equation}
\lb{ab4}
\d F \propto L^{\omega}
\end{equation}
where the exponent $\omega$ is known. On the other hand, if the typical deviation of the trajectory
from the origine is equal to $x$, then the loss of the energy due to the elastic term in the Hamiltonian
(\ref{a1}) must be of the order of $Jx^{2}/L$. Ballancing the two energies,
one can write the following estimate:

\begin{equation}
\lb{ab5}
\ol{\la x^{2}\ra} \sim \frac{L \; \d F}{J} \propto L^{\omega + 1}
\end{equation}
Then, according to the definition of the wandering exponent (\ref{ab3}), one finds the following simple
relation between the two exponents:

\begin{equation}
\lb{ab6}
2\z = \omega +1
\end{equation}

\subsection{The replica method}

The scaling of the free energy fluctuations with the size of the system $L$ can
be relatively easy investigated in terms of the replica method \cite{E,EA}.
To this end one has to calculate the average

\begin{equation}
\lb{ab7}
Z_{}(n) \equiv \ol{Z^{n}[v]}
\end{equation}
of the $n$-th power of the partition function:

\begin{equation}
\lb{ab2}
Z_{} [v] =  \int_{0<t<L}^{} Dx(t) \exp\left\{-\frac{H[x(t),v]}{T} \right\}
\end{equation}
obtained by the integration over all the trajectories with $x(0)=0$.

According to the definition of the free energy:

\begin{equation}
\lb{ab9}
F_{} = - T\ln Z_{}[v]
\end{equation}
the replica partition function $Z_{}(n)$ can be represented as follows:

\begin{equation}
\lb{ab10}
Z_{}(n) = \ol{ \exp\left[ -\frac{n}{T} F_{}\right]}
\end{equation}
The average in Eq. (\ref{ab10}) as everywhere above is calculated
over the realizations of random potential $v(x,t)$.
On the other hand the free energy $F\equiv F[v]$ is itself
the sample-dependent random quantity,
whose distribution function we shall denote as $P(F)$.
Then Eq. (\ref{ab10}) can be rewritten as

\begin{equation}
\lb{ab11}
Z_{}(n) = \int dF P(F) \exp\left[ -\frac{n}{T}F\right]
\end{equation}
which is nothing else but the Laplace transform of the free energy
distribution function $P(F)$.

The free energy corresponding to the replica partition function (\ref{ab7})
can be naturally defined as:

\begin{equation}
\lb{ab8}
F_{}(n) = -T\ln Z_{}(n)
\end{equation}
Although this quantity can be calculated only for integer $n$,
according to the standart ideology of the replica approach it
has to be considered as a function of the {\it continuous} parameter $n$ which
implies a necessity of an analytic continuation in $n$.

Let us represent the free energy $F_{}(n)$ of the replicated system
as a series in powers of the replica parameter $n$:

\begin{equation}
\lb{ab12}
F(n) = \sum_{k=1}^{\infty} \frac{F_k}{k!} n^{k}
\end{equation}
Then, taking the $k$-th derivative over $n$ at $n=0$ from both sides of the
Eq.(\ref{ab8}) for the $k$-th
order of the free energy fluctuations one finds:

\begin{equation}
\lb{ab13}
F_k=T^{1-k}\ol{\ol{ F^{k}}}
\end{equation}
where double overbar denotes the irreduceble average.

\subsection{The Bethe ansatz type solution}

In the framework of replica approach the statistical meachanics of the
{\em irregular} system is analyzed by considering the statistical mechanics
of the {\em regular} system in which the disorder manifests itself in the form
of the interaction between $n$ identical replicas of the original system.
For the system described by the Hamiltonian (\ref{a1}) and the Gaussian
statistics of the random potential, Eq.(\ref{a2}),
the averaging of $Z^n[v]$ over disorder leads to the expression for $Z(n)$
the form of which corresponds to the following replica Hamiltonian:

\begin{equation}
H_{repl}=\int_{0}^{L}dt\,\left\{\frac{J}{2T}\sum_{a=1}^{n}
\left(\frac{dx_a}{dt}\right)^2
-\frac{1}{T^2}\sum_{a,b=1}^{n}V[x_a(t)-x_b(t)]\right\}.           \label{a3}
\end{equation}

Eq. (\ref{a3}) has a form of the Eulidean (imaginary time) action
%The partition function corresponding to the Hamiltonian (\ref{a3})
%has a form of an imaginary time  functional integral
describing the quantum-mechanical system of $n$ particles with the mass $J/T$
and interaction $V(x)/T^2$. The same system can be described by the
quantum-mechanical (operator) Hamiltonian

\begin{equation}
\hat{H}=-\frac{T}{2J}\sum_{a=1}^{n}\nabla^2_a-\frac{1}{T^2}\sum_{a,b=1}^n
V(x_a-x_b)
                                                                 \label{a10}
\end{equation}
which for the classical partition function defined
by the Hamiltonion (\ref{a3}) plays the role of the transfer matrix.

In the limit of infinite size $(L\rightarrow\infty)$ the free energy of a
system (for any boundary conditions) is dominated by the highest eigenvalue of
transfer matrix or, in our case, by
the lowest eigenvalue $E_0$ of the quantum-mechanical
Hamiltonian (\ref{a10})
\begin{equation}
F(n)=TE_0(n)L                                                     \label{a4}
\end{equation}

For any integer $n$ the lowest eigen-value corresponds to the fully symmetric
(nodeless) wave-function which for the case of local correlations in $x$
direction

\begin{equation}
V(x) = u\delta(x)                                                 \label{a12}
\end{equation}
has been found exactly by Kardar \cite{K}:

\begin{equation}
\Psi_{0}[x_a]=\exp\left(-\mbox{\ae}\sum_{a,b=1}^{n}|x_a-x_b|\right)
                                                                \label{a13}
\end{equation} where for our choice of notation
\begin{equation}
\mbox{\ae}=\frac{Ju}{T^3}                                       \label{a14}
\end{equation}
The energy of this state is equal to

\begin{equation}
\lb{a15}
E_0(n) =  -\frac{V(0)}{T^2} n - \frac{Ju^2}{6T^5} n(n^{2} - 1)
\end{equation}
where the first term describes the trivial contribution to $E(n)$
related to the terms with $a=b$ in the second sum in Hamiltonian (\ref{a10}).

Substitution of Eq. (\ref{a15}) into Eq. (\ref{a4}) gives
\begin{equation}
\lb{a15a}
F(n) = F_{1} n + \frac{1}{6}F_{3} n^{3}
\end{equation}
where

\begin{equation}
\lb{a15b}
F_{1} = \left[-\frac{V(0)}{T} + \frac{Ju^2}{6T^4}\right]L
\end{equation}
and

\begin{equation}
\lb{a15c}
F_{3} = - \frac{Ju^2}{T^4}L
\end{equation}

Comparison of Eq. (\ref{ab13}) with Eq. (\ref{a15a}) shows that for
$L\rightarrow\infty$ the average free energy (per unit length) of a random
polymer

\begin{equation}
f\equiv\lim_{L\rightarrow\infty}\frac{{\overline{F(L)}}}{L}     \label{a16}
\end{equation}
is given by the linear in $n$ contribution to $F(n)$:

\begin{equation}
f=\frac{1}{L}\lim_{n\rightarrow 0}\frac{F(n)}{n}=\frac{F_{1}}{L} =
-\frac{V(0)}{T}+\frac{Ju^2}{6T^4}                            \label{a17}
\end{equation}
in which the first (formally divergent) term always dominates.
Therefore the average free energy of the system could be defined only after
proper short-scale regularization of the starting Hamiltonian.

However, the fluctuations of the free energy are quite well defined without
any regularization. According to Eqs. (\ref{ab13}) and (\ref{a15c})
the typical value of the free energy fluctuations can be estimated as:

\begin{equation}
\lb{a15d}
\d F \sim \left(\left|\,\overline{\overline{F^{3}}}\,\right|\right)^{1/3} =
\left(T^2 | F_{3}| \right)^{1/3} = \left(\frac{Ju^2}{T^2}\right)^{1/3}L^{1/3}
\end{equation}
Therefore, according to Eq. (\ref{ab5}) for the average
square deviation of the trajectory one finds the following result:

\begin{equation}
\lb{a15e}
\ol{\la x^{2}\ra} \sim \left(\frac{u}{JT}\right)^{2/3} L^{4/3}
\end{equation}

\sectio{Introduction of the regularization}

Apparently the divergence of the average free energy, Eq.(\ref{a17}), is removed if one takes
into account that in a
physical system correlations of random potential should be described by a
smooth function with a finite correlation radius
[and therefore a finite value of $V(0)$].
However in such case the quantum-mechanical problem defined by Eq. (\ref{a10})
cannot be solved exactly.
Nonetheless it seems reasonable to assume that for narrow enough $V(x)$ one
can still use the expression (\ref{a17}) in which now $u$ should stand for

\begin{equation}
u=\int_{-\infty}^{+\infty}dx\,V(x)                           \label{a18}
\end{equation}

It is easy to understand that such approximate description
[based on Eqs. (\ref{a13})-(\ref{a15c})] at low temperatures
has to fail. One has to remember that in case of the $\delta$-functional
interaction the wave function (\ref{a13})
is constructed as a generalization of two-particle problem wavefunction

\begin{equation}
\Psi(x_1,x_2)=\exp(-\mbox{\ae}|x_1-x_2|)                      \label{a22}
\end{equation}
On the other hand in the case of a rectangular well:

\begin{equation}
V(x)=\left\{\begin{array}{ll}
     V~~ & \mbox{for }|x|<r \\
     0   & \mbox{for }|x|>r \end{array} \right.               \label{a20}
\end{equation}
(for which $u=2Vr$) the wave function of the two-particle problem for
$|x_1-x_2|>r$ also has the form (\ref{a22}) with

\begin{equation}
\mbox{\ae}=\frac{1}{r}g\left(\frac{2T^3_0}{T^3}\right)        \label{a6}
\end{equation}
where

\begin{equation}
g(z)\approx\left\{\begin{array}{ll} z^{1/2}~~ & \mbox{for }z\ll 1 \\
                 z & \mbox{for } z\gg 1\end{array}\right.     \label{a7}
\end{equation}
and

\begin{equation}
T_0=(JVr^2)^{1/3}                                            \label{a21}
\end{equation}
It is not hard to check that the condition $\mbox{\ae}r\ll 1$ for this wave-function to be
wide in comparison with the well width coinsides with the condition $T\gg T_0$.
In such case the value of $\mbox{\ae}$ given by Eqs. (\ref{a6})-(\ref{a21})
coinsides with (\ref{a14}). In the opposite limit $T\ll T_0$ ($\mbox{\ae}r\gg 1$) the
two-particvle wave-function is almost completely localized inside the well
and cannot be used as a building block
for the construction of the solution of $n$-particle problem.

This gives a clear indication that for {\em arbitrary}
finite-width form of the function $V(x)$
describing the correlations of random potential the applicaton of the Kardar's
solution for the description of random polymer can work only at high enough
temperatures whereas in the low temperature limit the solution should be
different.

Quite paradoxically the differentiation of Eq. (\ref{a17}) shows that for $T\gg T_0$
the free energy defined by Eq. (\ref{a17}) corresponds to negative enthropy.
One should not be too scared of that property since in the approach discussed
above the free energy of a directed polymer is calibrated in such a way that
in the absence of the disorder it is equal to zero.
Therefore the total free energy
will be given by Eq. (\ref{a17}) {\em plus} the free energy in absence of
disorder. This second term will give the positive contribution to the
enthropy which will overcome the negative contribution from Eq. (\ref{a17}).

\sectio{The low temperature solution of the regularized problem}

Let us now consider the $n$-particle problem defined by Hamiltonian
(\ref{a10}) where

\begin{equation}
V(x)= V\left[1-b\frac{x^2}{r^2}\right]                      \label{b1}
\end{equation}
for $|x|<r$ and is equal to zero elsewhere.
Here in comparison with Eq. (\ref{a20}) we have introcuded a finite curvature
of the potential inside the well which is described by an additional free
parameter $b$ ($0<b\leq 1$). At some stage of calculation $b$ will be assumed
to be much smaller than one. For $V(x)$ of the form (\ref{b1})

\begin{equation}
u=2\left(1-\frac{b}{3}\right)Vr\sim 2Vr                     \label{b2}
\end{equation}
and therefore the characteristic temperature $T_0$ defining the range of
applicability of Kardar's approach ($T\gg T_0$) still can be chosen in
the form (\ref{a21}).

The characteristic frequency for small oscillations at the bottom
of such truncated parabolic well is given by
\begin{equation}
\Omega=\sqrt{\frac{2bV}{JTr^2}}                              \label{b3}
\end{equation}
and increases with decrease of $T$
much slower than the depth of the well $W=V/T^2$.
Thus the limit of low temperatures may correspond to the case
when all particles are localized near the bottom of the well.
In such limit the ground state energy $E(n)$ for the
$n$-particle system can be rather accurately found by assuming
that Eq. (\ref{b1}) holds for all $x_a-x_b$.

In such aproximation the ground state wave-function has a form

\begin{equation}
\Psi[x_a]=\exp\left[-\frac{1}{A\sqrt{2n}}
{\sum_{a,b=1}^n(x_a-x_b)^2}\right];~~
A=\frac{\Omega}{W}\frac{r^2}{b}
                                                            \label{b4a}
\end{equation}
whereas its energy is given by

\begin{equation}
E(n)=-Wn^2+\Omega\sqrt{\frac{n}{2}}(n-1)                    \label{b4b}
\end{equation}
(cf. with Refs. \cite{P} and \cite{M}).
In the following it will be convenient to keep in mind that the ratio
of $\Omega$ and $W$ can be expressed as

\begin{equation}
\frac{\Omega}{W}=\left(2b\frac{T^3}{T_0^3}\right)^{1/2}      \label{b5}
\end{equation}
It is not hard to find by a straightforward calculation that for
$\Psi(x)$ of the form (\ref{b4a})

\begin{equation}
\label{b6}
\langle(x_a-x_b)^2\rangle=\frac{A}{\sqrt{2n}}  = \fr{1}{\sqrt{2n}}\frac{\Omega}{W} \fr{r^{2}}{b}
\end{equation}
so for $\Omega/W\ll b$ (that is $T\ll b^{1/3}T_0$) $\Psi(x)$
is indeed nicely localized at the bottom of the well for any integer $n$
and all corrections to $E(n)$ due to non-parabolicity
can be only exponentially small.

On the other hand one can easily see why the limit $n\rightarrow 0$ is
dangerous. The width (\ref{b6}) of the wavefunction (\ref{b4a}) grows with
decrease in $n$ and becomes comparable with the width of the well $2r$
at $n\sim b^{-1}(T/T_0)^3$ and therefore for smaller $n$ the ground
state wave-function should have essentially different form.
The simplest way to let the particles enjoy their mutual attraction while
keeping their number in the well not too small consists in
splittting them into $n/k$ infinitely separated blocks of $k$ particles.
The energy $E(n,k)$ of such state with broken replica symmetry
is given by

\begin{equation}
E(n,k)=\frac{n}{k}E(k)=n\left[-Wk-\Omega\frac{1-k}{\sqrt{2k}}\right]
                                                             \label{b7}
\end{equation}
and has extremum (maximum) as a function of $k$.

Variation of the polymer free energy per unit length:

\begin{equation}
f(k)=T\left[-Wk-\Omega\frac{1-k}{\sqrt{2k}}\right]            \label{b7a}
\end{equation}
with respect to $k$ gives an equation for the position of the maximum:

\begin{equation}
-W+\Omega\frac{1+k}{(2k)^{3/2}}=0
                                                             \label{b7b}
\end{equation}
the solution of which for $\Omega/W\ll 1$ has a form

\begin{equation}
\label{b8}
k_*\approx\frac{1}{2}\left(\frac{\Omega}{W}\right)^{2/3} = \2 ( 2 b)^{1/3} \fr{T}{T_{0}}
\end{equation}
Substitution of Eq. (\ref{b8}) into Eq. (\ref{b6}) then shows that for

\begin{equation}
 T\ll b^{2/3} T_0                                               \label{b8a}
\end{equation}
the replicas belonging to the same block are indeed tightly bound to each other:

\begin{equation}
\langle(x_a-x_b)^2\rangle
\approx\left(\frac{\Omega}{W}\right)^{2/3}\frac{r^2}{b} \approx \fr{T}{b^{2/3} T_{0}} r^{2}  \ll  r^2
                                                                \label{b9}
\end{equation}
so the whole picture is really self-consistent.

Substitution of Eq. (\ref{b8}) into Eq. (\ref{b7a}) gives the temperature
independent expression:
\begin{equation}
f\approx-\frac{3}{2}(2b)^{1/3}\frac{V}{T_0}                     \label{b9b}
\end{equation}
which shows that in order to find the temperature dependence of $f$ in low
temperature limit we have to solve Eq. (\ref{b7b}) more accurately.
That gives
\begin{equation}
  k_*\approx \frac{1}{2}\left(\frac{\Omega}{W}\right)^{2/3}
            +\frac{1}{4}\left(\frac{\Omega}{W}\right)^{4/3}    \label{b9c}
\end{equation}
and
\begin{equation}
f\approx -\frac{3}{2}(2b)^{1/3}\frac{V}{T_0}
        +\frac{1}{4}(2b)^{2/3}\frac{V}{T_0^2}T                 \label{b9d}
\end{equation}

The idea to consider the state in which $n$ replicas are split into $n/k$
infinitely separated blocks of $k$ particles has been introduced by Parisi
\cite{P90}, who however applied it only to the case of local interaction
(\ref{a12}) ($\delta$-functional correlations in terms of original problem)
and discovered that free energy as a function of $k$ has extremum at $k=0$.
That is equivalent to considering all replicas belonging to the same block
right from the beginning.
Our analysis shows that smearing of the interaction potential leads
(at low enough temperatures)
to the shift of the extremum to non-trivial value of $k$ ($0<k<1$)
corresponding to splitting of replicas into blocks
(that is to replica symmetry breaking).

\vspace{5mm}

Although we have found that in the exremal solution the particles split
into $k$ separate blocks there are no reasons for these blocks to be
infinitely separated from each other. The presence of strong attraction
between the particles in each block makes it possible to consider such block
as a complex particle with the mass $kJ/T$, the interaction between these
complex particles being given by $k^2 V(x)/T^2$. The last expression
can be expected to be very accurate when we consider the temperature interval
(\ref{b8a}) in which the distances between the particles inside each block are
much smaller than the well radius $r$.

Therefore at low temperatures the behaviour of our system in which the
particles are assumed to be tightly bound in $n/k$ separate blocks
can be described by the Hamiltonian (\ref{a10}) in which

\begin{equation}
J\rightarrow kJ;~~V(x)\rightarrow k^2 V(x);~~u\rightarrow k^2 u \label{c1}
\end{equation}
Our earlier experience tells us that for some values of parameters such system
can be rather accurately described by the wave function of the form
(\ref{a13}) in which now $x_a$ stand for coordinates of different blocks.
The energy of such state will be given by Eq. (\ref{a15}) in which
substitutions (\ref{c1}) and $n\rightarrow n/k$ have to be made with the first
term being substituted by Eq. (\ref{b7}):
\begin{equation}
E(n,k)={n}\left\{-Wk-\Omega\frac{1-k}{\sqrt{2k}}
-2B\frac{W^3}{\Omega^2}k^4\left[\left(\frac{n}{k}\right)^2-1\right]\right\}
                                                              \label{c2a}
\end{equation}
where
\begin{equation}
B=\frac{2b}{3}\left(1-\frac{b}{3}\right)^2
                                                               \label{c2b}
\end{equation}
is a small parameter if $b$ is small.

All this leads to appearence in the expression for $f(k)$
of one more term [in comparison with Eq. (\ref{b7a})]:

\begin{equation}
f(k)=T\left[-Wk-\Omega\frac{1-k}{\sqrt{2k}}+2B\frac{W^3}{\Omega^2}k^4\right]
                                                               \label{c2}
\end{equation}
which describes the contribution related to mutual interaction between the
blocks. Substitution of Eq. (\ref{b8}) into the saddle-point equation

\begin{equation}
-W+\Omega\frac{1+k}{(2k)^{3/2}}+8B\frac{W^3}{\Omega^2}k^3=0
                                                             \label{c3}
\end{equation}
which is obtained by variation of Eq. (\ref{c2}) shows that for $B\ll 1$
(that is for $b\ll 1$)
the maximum of $f(k)$ still exists and in the lowest order in $b$ the position
of this maximum is still given by Eq. (\ref{b8}).

The applicability of such approach requires that the distances between the
blocks should be much larger than the size of the well (exactly in same way as
when Kardar type solution is constructed from the separate particles and not
from the blocks):
\begin{equation}
\mbox{\ae}r\ll 1                                                 \label{c5}
\end{equation}
Sustitution of Eqs. (\ref{a14}), (\ref{c1}) and (\ref{b8}) into Eq. (\ref{c5})
then reduces it to condition
\begin{equation}
\frac{b}{2}\left(1-\frac{b}{3}\right)\ll 1
\label{c6} \end{equation}
which apparently is equivalent to the same condition $b\ll 1$.

Strictly speaking the expression for $f(k)$ given by Eq. (\ref{c2}) has also
another extremum (minimum) at
\begin{equation}
k=k_{**}\approx\frac{1}{B^{1/3}}k_*                              \label{c7}
\end{equation}
but analogous analysis shows that
\begin{equation}
\mbox{\ae}(k_{**})r\approx\frac{3}{4}\left(1-\frac{b}{3}\right)^{-1}\sim 1
                                                                 \label{c8}
\end{equation}
and therfore this second extremum takes place in the domain of parameters
where expression (\ref{c2}) based on the assumption (\ref{c5}) can no longer
be trusted.

\vspace{5mm}

The form of Eq. (\ref{c2a}) shows that in the considered case the only
non-linear (in $n$) contribution to $E(n)$ is also of the third order in $n$
and corresponds to

\begin{equation}
\lb{c9}
F_{3} = - \fr{4V T_{0}^{3}}{ T^{4}} k^{2}L
\end{equation}
where we assume $b\ll 1$.
Using the saddle-point value of the parameter $k$ one gets:

\begin{equation}
\lb{c10}
F_{3}(k_*) = -(2 b)^{2/3}\fr{VT_0}{T^{2}}L
\end{equation}
Correspondingly, for the typical value of the free energy fluctuations one obtaines
the following temperature independent result:

\begin{equation}
\lb{c11}
\d F \sim \left( T^{2} |F_{3}| \right)^{1/3} L^{1/3} =C _{0} L^{1/3}
\end{equation}
where

\begin{equation}
\lb{c12}
C_{0} = (2 b)^{2/9} (V T_{0})^{1/3}
\end{equation}
Finally, for the mean square deviation of the polymer trajectory one
again finds $\ol{\la x^{2}\ra} \sim a  L^{4/3}$
with the temperature independent prefactor
\begin{equation}
a=\frac{C_0}{J}=\left(2b\frac{V^2r}{J^4}\right)^{2/9}            \label{c14}
\end{equation}

\sectio{Conclusion}

Thus we have demonstrated that in the case when random potential correlations
are characterized by finite correlation radius $r$
the solution at low temperatures has
essentially different structure than at high temperatures.
Nonetheless the value of the wandering exponent in both cases is the
same: $\z=2/3$ .
In contrast to the high temperature limit for which the prefactor $a$ in
scaling law (\ref{ab3}) is temperature dependent: $a\propto T^{-2/3}$,
in the low temperature limit it saturates at finite value $a\propto r^{2/9}$.

Since the value of the wandering exponent for both regimes is the same
there are no reasons for the sharp transition between these regimes.
At very low temperatures the solution is characterized by the one-step
replica symmetry breaking that is the replicas are splitted into $k$
well separated blocks.
With growth of temperature the distance between the blocks becomes
comparable with the size of each block. At higher temperatures the replica
symmetry breaking phenomena can manifest itself only in slight modulation of
the distance between nearest replicas in comparison with what follows from the
"replica symmetric" wave-function (\ref{a13}).

\sectio{Acknowledgements}

V.D. would like to thank M.Mezard for stimulating discussions.

\end{document}